# Non-invasive light observer

F. Morichetti[1*], S. Grillanda[1], M. Carminati[1], G. Ferrari[1], M. Sampietro[1], M. J. Strain[2], M. Sorel[2], A. Melloni[1]

[1]Dipartimento di Elettronica, Informazione e Bioingegneria, Politecnico di Milano, 20133 Milano, Italy.

[2]School of Engineering, University of Glasgow, Glasgow, G12 8LT, UK.

*Correspondence to: francesco.morichetti@polimi.it

**Photonic technologies lack non-invasive monitoring tools to inspect the light inside optical waveguides. This is one of the main barriers to large scale of integration, even though photonic platforms are potentially ready to host several thousands of elements on a single chip[1,2]. Here, we demonstrate non-invasive light observation in silicon photonics devices by exploiting photon interaction with intra-gap energy states localized at the waveguide surface. Light intensity is measured through a ContactLess Integrated Photonic Probe (CLIPP) that introduces no measurable extra-photon absorption and a phase perturbation as low as 0.2 mrad, comparable to thermal fluctuations of less than 3 mK. Multipoint light monitoring is demonstrated with a sensitivity of -30 dBm and a dynamic range of 40 dB. CLIPP technology is simple, inherently CMOS compatible, and scalable to hundreds of probing points per chip. This concept provides a viable way to real-time conditioning and feedback control of densely-integrated photonic systems.**

The concept of non-invasive light observation is still an open issue, if not a paradox. Optical power detection in photonic integrated circuits is typically realized by tapping a fraction of the light from the waveguide and rerouting it to a photodetector[3,4]. Alternatively, a short section of the waveguide can be used itself as a photodetector. In silicon waveguides, where material absorption is inhibited at wavelengths above 1.1 μm, photocarrier generation has been demonstrated by exploiting intrinsic sub-bandgap absorption mechanisms, including two photon absorption (TPA)[5,6], surface-state absorption (SSA)[7,8], and defect mediated absorption induced via selective ion-implantation[9,10]. Optical detectors proposed so far based on these effects require the waveguide core to be electrically contacted through highly doped regions and/or silicon/metallic electric lines to sweep out carriers efficiently from the absorbing region, thus leading to an intrinsic perturbation of the optical field. Non destructive photon observation has been successfully demonstrated in quantum optics[11,12], but quantum non demolition approaches appear hardly portable onto densely integrated photonic systems.

The physical effect exploited in our technique is the change of the waveguide conductance induced by native interaction of photons with the intra-gap energy states localized at the Si-SiO$_2$ interface[13] that exist even in an ideal roughness-free interface[14]. A capacitive access to the waveguide is used, thereby avoiding direct contact with the waveguide core, and no specific treatments at the waveguide surface need to be done. Neither photon spilling is required nor appreciable perturbation on the optical field is introduced, thereby implementing a truly non-invasive light observation.

As shown in Fig. 1a, the CLIPP simply consists of two electrodes realized onto an electrically insulating layer acting as upper cladding (Fig.1b). Owing to the doping level of commercial SOI wafers ($N_h$ = 10$^{15}$ cm$^{-3}$ p-type), in the electric domain the waveguide core behaves like a resistor $R_{WG}$, with about $10^2$ free holes distributed in the volume of a 1-µm-long waveguide section. For the channel waveguide of Fig. 1b (details on technology and design in the Supplementary Information), the electrodes are spaced from the silicon core ($w$ = 480 nm, $h$ = 220 nm) by a 1-µm-thick silica film, providing the access capacitance $C_A$ to the waveguide. At the wavelength of 1550 nm, the optical intensity at the $SiO_2$-metal interface is 80 dB below its peak value in the waveguide, so that the electrodes do not induce any detectable change in the waveguide propagation loss, amounting to 2 dB/cm. The light-induced conductance variation $\Delta G = \Delta(1/R_{WG})$, due to free carrier generation at the waveguide surface (Fig. 1c), is electrically accessed through the capacitance $C_A$ and monitored through an impedimetric measurement system (Fig. S1) with a sensitivity of 80 pS in conductance measurement[15] (Methods).

The CLIPP effectiveness in measuring the light dependent conductance variation $\Delta G$ is shown in Fig. 2 versus the local optical power $P$ for a quasi-transverse electric (TE) polarized light. The inset shows a top-view photograph of the device with a 100-µm-long waveguide section enclosed between two narrow electrodes. Light intensity was observed over a dynamic range of 4 orders of magnitude, down to -30 dBm (1 µW), this performance largely fulfilling the requirements of most practical applications. A response time down to 50 µs was achieved by driving the device at a frequency $f_e \sim$ 1 MHz and at a voltage $V_e \sim$ 1 V (Fig. S2). Although photons interaction with the surface states depends on the waveguide shape[7], because of the different mode overlap modes with the waveguide boundaries, single mode ($w$ = 480) and multi mode ($w$ = 1 µm) waveguides exhibit very similar performance. This small sensitivity implies that the applicability of the technique does not impose constraints to the design of the optical waveguide.

An in-depth analysis was carried out to identify SSA as the physical effect responsible for the change of waveguide conductance $\Delta G$. Results show that $\Delta G$ is associated with a carrier density variation rather than with a carrier mobility variation, and that TPA–mediated photocarrier generation is negligible (Supplementary Information, Figs. S3 and S4). The $\Delta G$-$P$ curves show a sub-linear behavior, where the slope is related to the waveguide perimeter with respect to the cross-sectional area and the intercept depends on the local density of defect states. As observed also in metal oxide semiconductor (MOS) devices[16] and in high gap conductors[17], this sub-linear relationship is consistent with a situation in which the number of photo-generated carriers is larger than those thermally available, so that an increase in the optical power results not only in a larger carrier density but also in a reduction of their recombination time[18]. This condition is verified in both devices of Fig. 2, where, at $P = $ -10 dBm ($\Delta G \approx$ 5 nS), about $10^2$ $\mu m^{-1}$ free carriers are locally photogenerated at the surface of a 1-µm-long waveguide section, this number being comparable with the number of native free holes distributed across the waveguide in absence of light.

Many light observers can be placed in strategic positions around the chip enabling the real time management of complex integrated optical systems. We fabricated a racetrack silicon resonator (Fig. 3a) equipped with three CLIPPs, one inside the resonator and the other two at the Through and Drop ports of the resonator, respectively. The same metal technology of the CLIPP is used to realize a thermo-optic actuator inside the resonator for the feedback control of its resonant wavelength[19]. To set a benchmark for performance evaluation, the CLIPPs were first switched off ($V_e = 0$) and the reference transmission spectra of the resonator were measured (Fig. 3b and 3c, purple curves) through a high precision tunable laser, with 1 pm resolution, synchronized with an optical spectrum analyzer (OSA). The measured 3 dB linewidth and free spectral range are 36 pm and 860 pm, respectively, resulting in a Q factor of about 43000. Measurements were repeated with the three probes sequentially switched on at $V_e$ up to 10V, and no detectable changes of the resonator spectrum and of the intrinsic Q factor (about 55000) were observed. The transmission spectra were then measured by using the CLIPPs at the Through (Fig. 3b) and at the Drop (Fig. 3c) ports. The electric signal of the CLIPP provides a direct measurement of the resonance wavelength of the resonator, with less than 1 pm shift compared to the OSA reference. By using the calibration curve of Fig. 2, we also retrieved the spectral line and the optical power level inside the resonator, as shown in Fig. 3d. The 3 dB linewidth, measured at a power level of -30 dBm, is less than 2 pm larger than the OSA measurement, thus demonstrating the accuracy of the CLIPP.

The non-invasive nature of the CLIPP was explored by looking for the existence of a signal component at the frequency $f_e$ in the output optical signal with a highly sensitive lock-in based detection system (Methods). As no evidence of disturbance was detected on straight waveguides, we investigated the perturbative effects on high-Q optical microresonators. We injected a continuous-wave (CW) light into the resonator of Fig. 3a and measured the ratio between the component $P(f_e)$ at $f_e = 2$ MHz and the average power $P_{CW}$ of the optical signal outgoing from the Through port (Fig. 4a, purple curve). The inner CLIPP was driven at $V_e = 1$ V and the wavelength scanned 0.2-nm around the resonant wavelength $\lambda_r = 1555.889$ nm. The relative perturbation $P(f_e)/P_{CW}$ is minimum at $\lambda_r$, where the slope of the transmission spectrum vanishes, whereas it exhibits local maxima at $\lambda_p = \lambda_r \pm 9$ pm, where the slope is maximum. This result provides clear evidence that the perturbation is due to a refractive index modulation effect, causing a tiny wavelength shift of the resonator spectrum and an intensity modulation of the output signal[20]. The estimated wavelength shift is about 55 fm (6.6 MHz), that is less than 0.1% of the resonator linewidth. This makes the CLIPP applicable without significant disturbance even to the silicon microring resonators with the highest Q factor ever realized, which is in the order of $8\times10^5$ (Supplementary Information)[21]. Figure 4b shows that the associated phase modulation is as low as 0.2 mrad for $V_e = 1$ V, corresponding to about 0.2 ppm of the effective index of the waveguide (Fig. S5). It should be noted that such perturbation is comparable to that induced by a temperature fluctuation of 3 mK, which is at least two orders of magnitude below the thermal stability limit guaranteed by conventional thermo electric coolers and can be considered negligible in practical applications. The origin of the measured perturbation is addressable to a linear electro-optic effect; we estimated a small second order susceptibility $\chi^{(2)}$ of about 5 pm/V, that is in line with previously reported results for non intentionally strained silicon waveguides[22].

The CLIPP enables direct monitoring of a generic photonic device embedded inside an arbitrary complex architecture instead of inferring its behavior from global output signals, as presently imposed, thus considerably reducing the complexity of identification algorithms. Scalability to very many CLIPPs per chip, enabling multi-point monitoring and feedback controls in a photonic chip, is conceivable thanks to the compactness of the CLIPP structure and to the possibility to integrate electronic read-out and processing circuits with hundreds of amplifiers and mixers in a lock-in architecture[23]. Thanks to the compatibility with complementary metal oxide (CMOS) production lines, direct integration of the electronic circuitry into the same photonic chip is also feasible. These considerations provide a viable way to make photonics break away from today's device level up to a system-on-chip level1, so as to fulfill the requirements of wide-ranging application fields, such as telecom[24], optical interconnects[25], biosensing[26], quantum manipulation and computing[27,28,29], and chip-to-chip quantum communications[30]. Furthermore, the possibility of pasting a CLIPP in any point across a photonic wafer makes it a powerful tool for wafer scale automated testing, without requiring expensive bulk optical measurement rigs.

**Methods Summary**

**Experimental setup for the measurement of the waveguide impedance.** The light is coupled to the silicon waveguides through micro-lensed tapered fibers (1.7 µm mode field diameter) positioned by means of piezo-actuated translational stages with 5 nm closed-loop accuracy. The temperature of the optical chip is controlled within 0.1 K by a Peltier thermocooler integrated inside a customized holder. The state of polarization of the input light is controlled with a polarization controller providing an extinction ratio greater than 30 dB between the excited quasi-TE and quasi-TM modes over a wavelength range of 20 nm around 1550 nm.

The real and imaginary parts of the waveguide impedance are measured through a synchronous detection architecture, including a low-noise transimpedance amplifier coupled to a high precision lock-in demodulator. To measure $\Delta G$, a sinusoidal voltage $V_e$ at a frequency $f_e$ is applied to one electrode of the CLIPP, the other being connected to a transimpedance amplifier to sense the flowing current $i_e$. The transimpedance amplifier has a $10^4$ V/A conversion gain, a bandwidth of 80 MHz and a 6 pA/√Hz minimum noise floor at 1 Hz integration bandwidth, resulting in a total measured rms noise of ~80 pS. The measured wavelength scans of the ring resonator (Fig. 3) are performed at a rate of 8 pm/s. In the perturbation measurement of Fig. 4, the output optical signal is detected by an external photodiode and sent to a second channel of the lock-in detector, enabling synchronous detection of the optical and the electrical signals.

**Acknowledgments**

We thank A. Canciamilla for the support in the design of the photonic circuits, D. Bianchi for discussions on the electronic instrumentation design, and the James Watt Nanofabrication Centre (JWNC) staff at Glasgow University for the fabrication of the devices. This work was supported by the Italian PRIN 2009 project Shared Access Platform to Photonic Integrated Resources (SAPPHIRE), by Fondazione Cariplo, (grant n.2011-2118), and European Project BBOI of the 7th EU Framework Program.

**Author Contributions** F. M., A. M. and M. Sampietro developed the CLIPP concept and designed experiments. S. G. and M. C. designed the integrated circuits, performed experiments and collected experimental data. F. M., S. G., M. C. and G. F. performed the analysis of the experimental data. M. J. S. and M. Sorel fabricated the samples. A. M. and M. Sampietro supervised the project. All authors discussed the results and implications, and contributed to the preparation of the manuscript.

The authors declare no competing financial interest. Correspondence and requests for materials should be addressed to F.M. (francesco.morichetti@polimi.it).

**Figure 1**

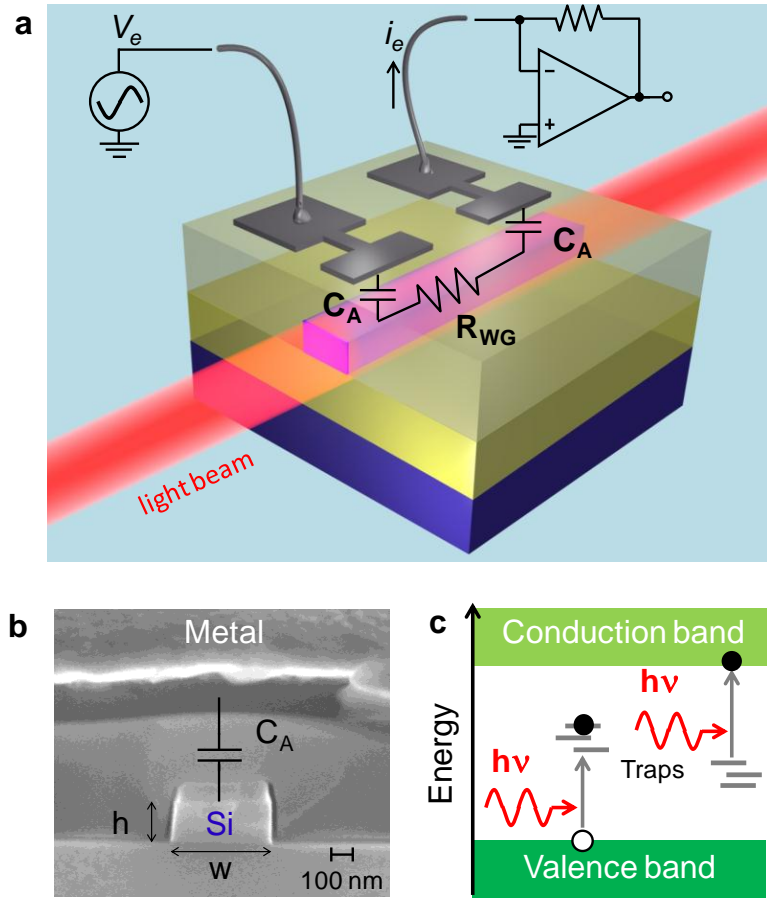

**Figure 1 | Non-invasive light observer integrated on a silicon chip. a,** Illustration of the device consisting of two metal electrodes deposited onto the electrically-insulating upper cladding and capacitively coupled ($C_A$) to the electrically-resistive ($R_{WG}$) silicon nanowaveguide. **b,** SEM photograph of the silicon waveguide cross section in the CLIPP section. **c,** Sketch of two possible alternative SSA mechanisms implying intra-gap energy states concurring to the creation of a free carrier and a corresponding recombination center.

**Figure 2**

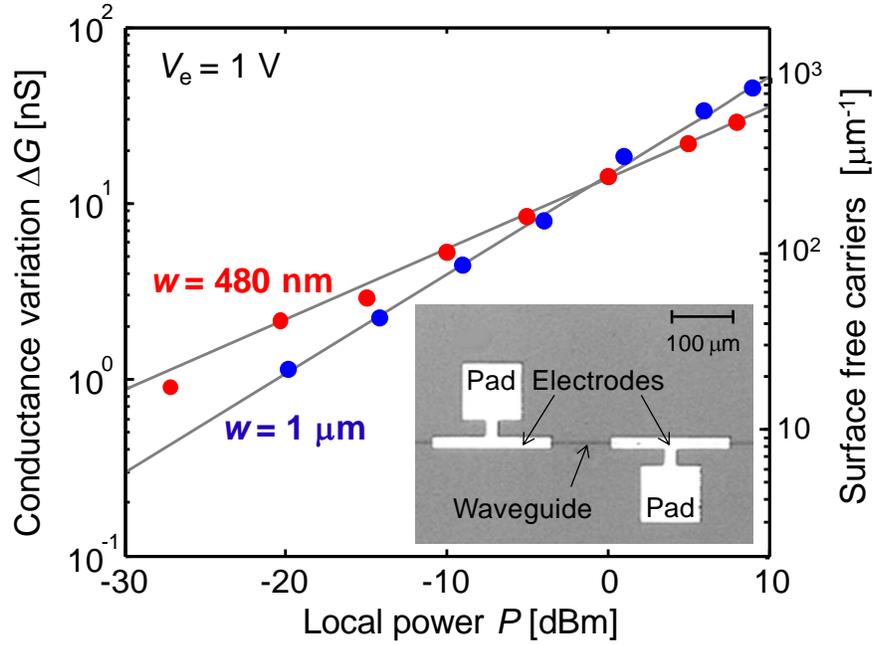

**Figure 2 | Optical power monitoring performed through a CLIPP.** The conductance variation $\Delta G$ of silicon waveguides versus the optical power $P$ is detected by impedance spectroscopy technique ($V_e = 1$ V) through two metal electrodes (20 μm × 200 μm) at a distance of 100 μm. The 100 μm × 100 μm metal pads are used for the wire bonding with the impedimetric read-out system. A single mode ($w = 480$ nm, red circles) and a multimode (w = 1 μm, blue circles) silicon waveguide, fabricated in two different runs, demonstrate the flexibility of the CLIPP approach.

**Figure 3**

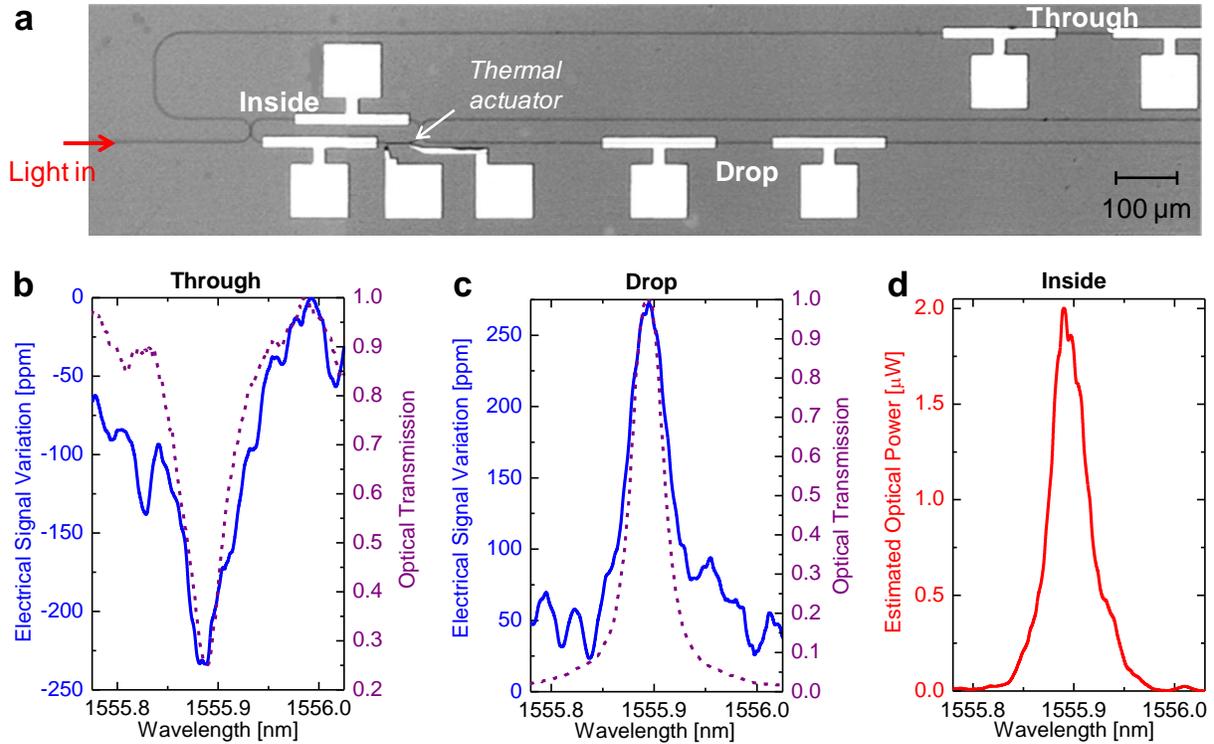

**Figure 3 | Non-invasive monitoring of a silicon photonic integrated circuit. a,** Top-view optical microscope photograph of a racetrack microresonator with a CLIPP at both the Through and the Drop port bus waveguides, and with an inner CLIPP inside the resonator. The resonator has a radius of 20 μm and a geometric length of 644 μm. **b,** Through and **c,** Drop port TE transmission measured with an external OSA (purple dashed curves) and with the on-chip CLIPPs (blue solid curves). The amplitude and the frequency of the electric signal driving the CLIPPs are $V_e = 1$ V and $f_e = 2$ MHz, respectively. **d,** The intracavity optical power around the resonant wavelength is retrieved from the electric signal of the CLIPP inside the resonator.

**Figure 4**

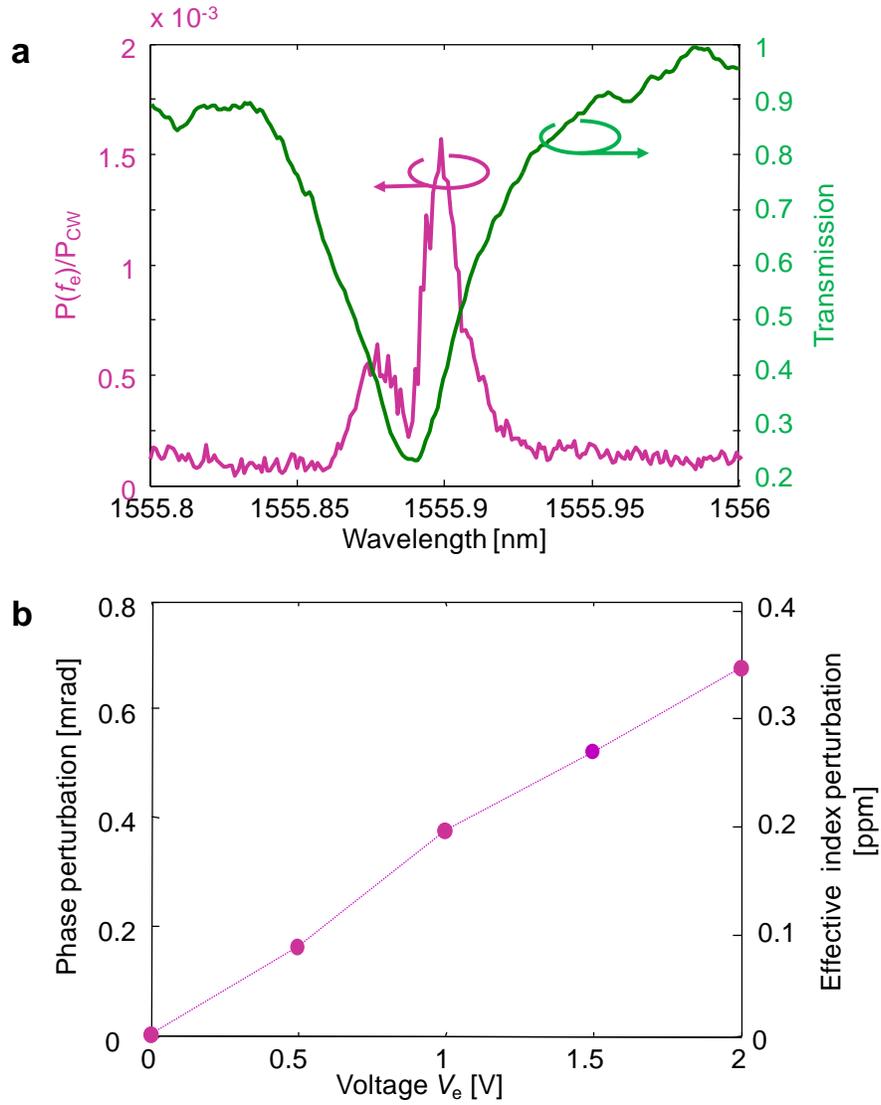

**Figure 4 | Analysis of the perturbative effects induced by the CLIPP. a,** The purple curve shows the ratio between the perturbation $P(f_e)$, given by the power of the optical signal modulated at the frequency of the electric signal, and the power of the optical signal $P_{CW}$. The measurement is carried out on the optical signal outgoing from the Through port of the resonator of Fig. 3. The asymmetry in the measured perturbation is inherently associated to the non-symmetric transmission spectrum of the resonator (green curve). **b,** Phase perturbation and effective index perturbation induced by the CLIPP.

**Supplementary Information**

We provide in the following additional details and experimental data on our demonstration of a non-invasive light observer by supplementing optical and electrical characterization, experimental setup description and fabrication methods.

*Device structure and fabrication of the CLIPP*

The silicon waveguides are fabricated on a commercial silicon-on-insulator (SOI) wafer with a 220-nm thick silicon layer on a 2-μm thick oxide buffer layer. The waveguide pattern is written on a hydrogen silsesquioxane (HSQ) resist through electron-beam lithography and then transferred to the silicon core by an inductively coupled plasma etching process[31]. A residual 80-nm thick HSQ cap layer on top of the waveguide is visible in the SEM picture of Fig. 1b. The waveguide core is buried under a 1-μm thick cover layer, consisting of 550 nm of spun and baked HSQ and 450 nm of silicon dioxide grown by plasma enhanced chemical vapor deposition (PECVD). Optical coupling loss with optical fibers was reduced down to 5 dB/facet through modal adapters realized at the chip's facets by an inversely tapered section of the silicon waveguide, buried in a SU-8 polymer waveguide, which also reduces the spurious Fabry-Perot resonance effects due to optical reflections at the end sections of the waveguide. The electrodes of the CLIPP consist of a 200 nm thick Au film deposited onto the silica cladding (with an intermediate 20 nm thick Ti adhesion layer) and patterned by a lift-off technique. It should be noted that the CLIPP can be fabricated by using any CMOS compatible metal technology, and that can exploit conventional technologies used for realizing thermal actuators, like those of Fig. 3a, without any additional process steps[19].

*Device impedance modeling and electrode design*

An accurate modeling of the silicon waveguide electric impedance is essential for the design of the measurement system and the correct interpretation of the experimental data. The simplified equivalent model is illustrated in Fig. S1. $C_A$ is the access capacitance and its value can be estimated (neglecting fringing field effects) as a parallel plate capacitor $C_A = \varepsilon_0 \cdot \varepsilon_{ox} \cdot w \cdot L_e / t_{ox}$ where $w$ is width of the waveguide, $L_e$ the length of the metal pad and $t_{ox}$ (1 µm) and $\varepsilon_{ox}$ (3.9) are respectively the thickness and relative dielectric constant of the SiO$_2$ top layer. Similarly, the nominal value of the waveguide resistance in absence of light is $R_{WG} = \rho \cdot L / (w \cdot h)$ where $\rho$ is the bulk resistivity of the waveguide (~1 MΩ/µm), $L$ the distance between the electrodes, and $h$ the waveguide thickness (220 nm). In parallel to the main $C_A$-$R_{WG}$ path to be probed, there is a parasitic capacitance $C_E$ due to the stray coupling between the electrodes (measured in the 0.16 – 0.8 pF range, depending on the position of the connection wires). Since the impedance $C_E$ is about 3 orders of magnitude smaller than $R_{WG}$ in the MHz range, a 100 ppm resolution lock-in is required.

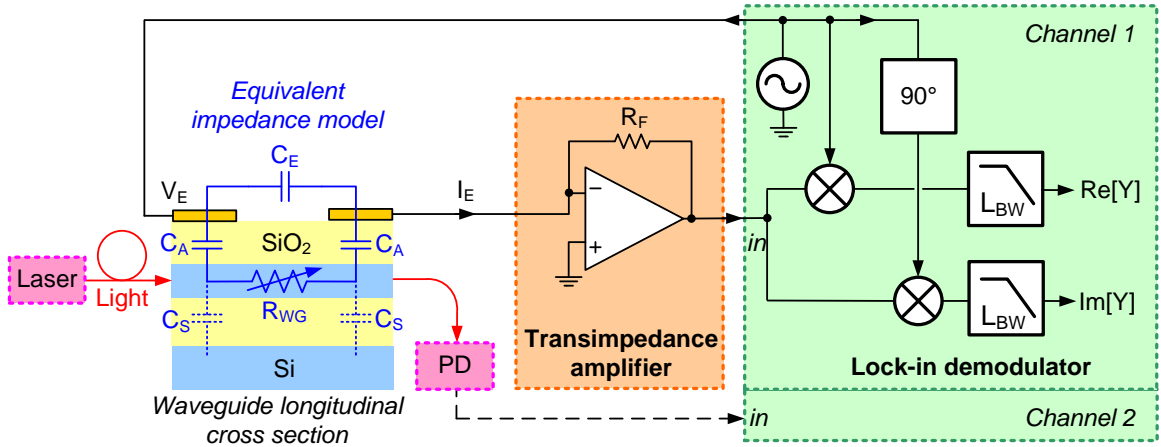

**Figure S1.** Scheme of the experimental setup adopted for admittance measurements. The equivalent lumped-parameter impedance model of the device is shown on the longitudinal cross section of the waveguide. A low-noise transimpedance amplifier is combined with a precision lock-in detector.

As illustrated in Fig. S2, the change of waveguide conductance $\Delta G$ can be extracted by subtracting the reference spectrum of the waveguide admittance $Y_{WG}= 1/G_{WG}$ (measured in absence of light) from the admittance spectra measured at different optical power levels $P$. In order to correctly measure $\Delta G$, the resistive plateau visible in Fig. S2 must be located within the measuring range of the detection instrumentation (50MHz), i.e. the access pole frequency $f_p = 1/(2\pi \cdot C_A/2 \cdot R_{WG}) < 5$MHz. Since $\rho$, $w$, $h$, $t_{ox}$ are chosen independently, based both on fabrication constraints and optical optimization criteria, the degrees of freedom in the CLIPP design are the sizing of $L$ and $L_e$. By setting $L = 100$ µm ($C_A \sim 5$fF) and $L_e = 200$ µm, the resulting $f_e$ is about ~1 MHz. Thus, given the measured spectra, $f_e$ has been set to 2 MHz. Further miniaturization (reduction of $L$ and $L_e$) is possible at the price of operating at higher $f_e$ (since $f_p$ shifts at higher frequencies reducing $L$ or $L_e$). However, a physical upper limit to $f_e$ is set also by the presence of the stray capacitances $C_S$ between the waveguide and the silicon substrate through the bottom oxide (BOX, thickness 2 µm). In fact, this distributed capacitance attenuates the applied signal due to the high frequency partition between $C_A$ and $C_S$.

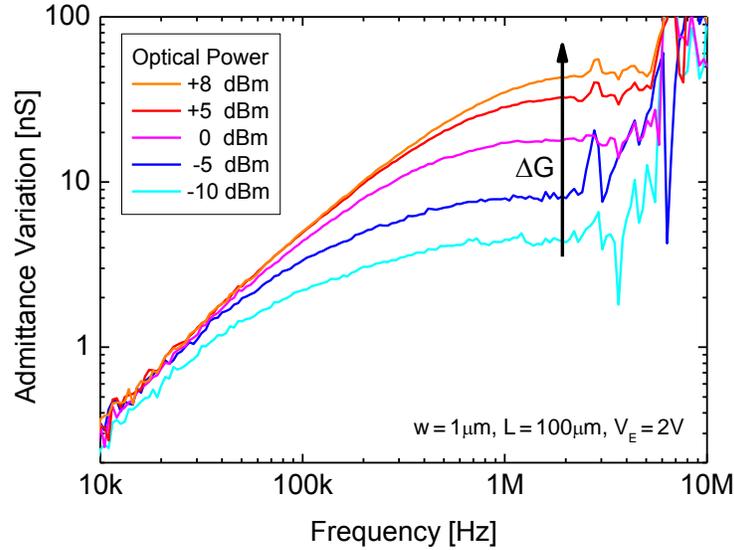

**Figure S2.** Measured differential admittance spectra (with respect to the absence of light) for increasing optical power levels, from which the variation of conductance $\Delta G$ is extracted**.**

The operative frequency $f_e$ slightly depends on the thickness $t_{ox}$ of the oxide layers through the access capacitance $C_A$. Standard deposition techniques guarantee a control of better than 10 nm in the thickness of the waveguide uppercladding. In our device $t_{ox} = 1$ μm and this small uncertainty introduces no significant variability in the CLIPP performance, because it only implies a 1% change of the pole frequency $f_p$, and thus of the operative frequency $f_e$ of the electrical detection (Fig. S2).

*Analysis of the physical effects affecting the waveguide electric conductance*

In this section we identify surface-state absorption (SSA) as the physical effect responsible for the change of waveguide conductance $\Delta G$ versus the local optical power $P$ shown in Fig. 2. The waveguide conductance $G$ increases with the geometrical waveguide cross section $A = wh$ and decreases with the length $L$ of the CLIPP according to the relation

$$G = \sigma \frac{A}{L}, \tag{s3}$$

where the electrical conductivity

$$\sigma = q(\mu_e N_e + \mu_h N_h) \tag{s4}$$

of the material depends on the electric charge $q$, on the mobility of electrons ($\mu_e$) and holes ($\mu_h$), and on the volume density of electrons ($N_e$) and holes ($N_h$). Since $A$ and $L$ are fixed by the waveguide geometry, the measured $\Delta G$ is essentially due to a conductivity variation $\Delta\sigma$, that is to either a variation in the mobility ($\Delta\mu$) and/or in the density ($\Delta N$) of electrons and holes.

Here, we first demonstrate that a variation $\Delta\mu$ in the carrier mobility does not explain the measured $\Delta\sigma$. Then, focusing on carrier density variation $\Delta N$, we also rule out the existence of significant photocarrier generation mediated by two-photon absorption (TPA) effects, proving that surface state absorption (SSA) is the only candidate to be responsible for the observed changes in the waveguide conductance.

*Variations of carrier mobility ($\Delta\mu$).* Typically in Si optical waveguides the mobility of carriers varies in the presence of: (a) radiation pressure and electrostriction induced optical forces, (b) temperature and (c) carrier density variations.

(a) *Radiation pressure and electrostriction induced optical forces.* From the measured $\Delta G$ versus $P$ (see Fig. 2) we infer the value of $\Delta\mu$ that would be required to justify the observed change in the waveguide conductance. By substituting Eq. (s4) in Eq. (s3), and assuming that in the p-doped silicon ($N_h = 10^{15}$ cm$^{-3}$) the conductivity is essentially due to free holes, we obtain

$$\Delta\mu = \frac{\Delta\sigma}{2qN_h} = \frac{1}{2qN_h}\frac{L}{A}\Delta G. \tag{s5}$$

Figure S3 shows the mobility variation for the same optical power range of Fig. 2, having considered a carrier density of $10^{15}$ cm$^{-3}$, typical of commercial SOI wafers. For instance, when the 480 nm wide waveguide operates at $P = 0$ dBm, we calculate $\Delta\mu = 417$ cm$^2$/(V·s), which corresponds to a change in the relative hole mobility of $\Delta\mu/\mu_h = 0.93$ (assuming a typical value $\mu_e = 450 cm^2/V\cdot s$ for the mobility of the free holes). Such mobility variation would require the application of a pressure to the waveguide of about $10^7$-$10^8$ Pa[32]. However, when operating in the optical power range of Fig. 2 (from -30 to 10 dBm) the typical radiation pressure level on the waveguide is lower than $2\cdot 10^3$ Pa[33]. Therefore, we rule out radiation pressure induced optical forces as the physical effect responsible for the observed photocarrier generation. Likewise, we exclude electrostriction, because it induces optical forces of about the same order of magnitude as radiation pressure[34].

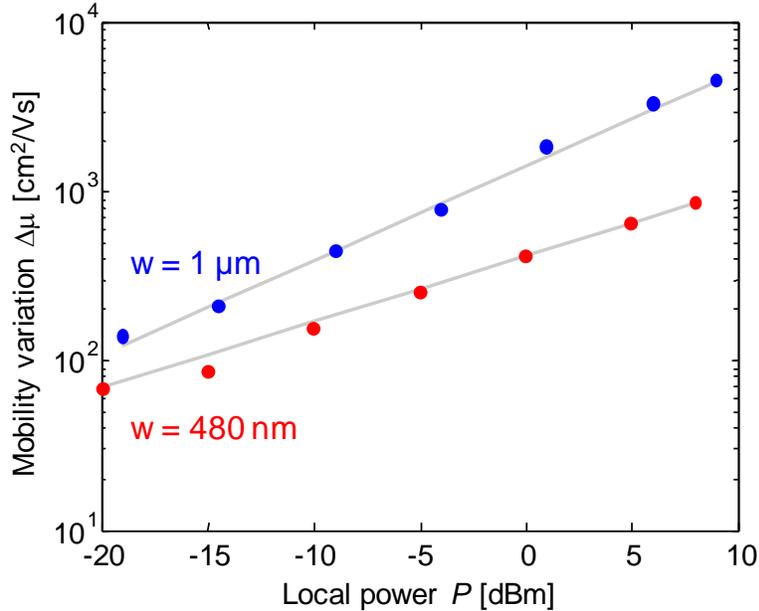

**Figure S3.** Variation of the carriers mobility as a function of the local optical power for the waveguides with width of 480 nm (red circles) and 1 μm (blue circles); the grey solid line provides a linear fitting of the experimental points.

(b) *Temperature variations.* The change of the waveguide conductance $\Delta G$ versus the waveguide temperature was measured in absence of light by controlling the chip temperature through a thermo-optic cooler placed below the sample. Figure S4 shows that 1 °C temperature change is responsible for $\Delta G = 1.17$ nS, this implying that a temperature variation by tens of degrees would be required to justify the results of Fig. 2. However, in this case we should observe the effects of this large temperature change also in the optical domain, as a wavelength shift of the waveguide transmission spectrum, amounting to 70 pm/°C for the single mode silicon waveguide with $w$ = 480 nm[35]. However, we do not observe any significant wavelength shifts (> 1 pm) for the optical power range employed in the CLIPP measurement.

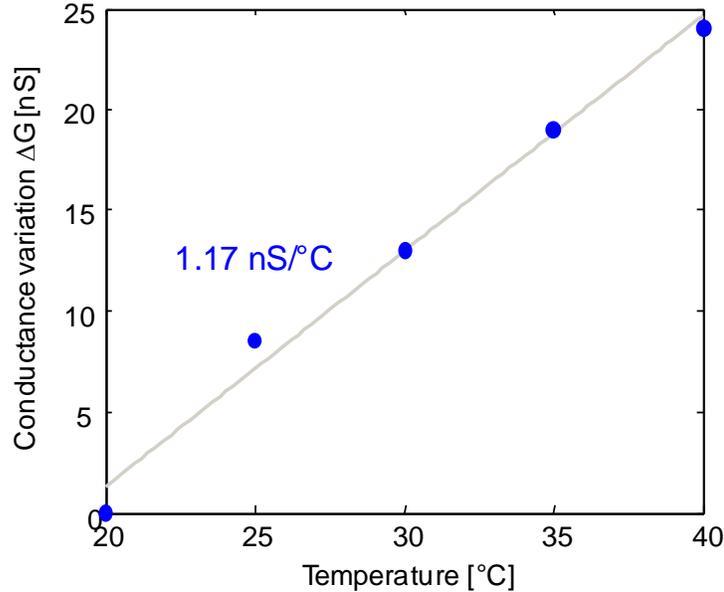

**Figure S4.** Measured conductance variation in absence of light in the waveguide CLIPP with a width of 1 µm as a function of temperature (blue circles); the grey solid line provides a linear fitting of the experimental points.

*(c) Carrier density variations.* In principle, a variation in the density of carriers induces a change in their mobility[36,37]. At a typical doping level of $10^{15}$ cm$^{-3}$ for SOI wafers, to halve the carrier mobility it is necessary to increase their density by at least two orders of magnitude[36,37]. However, in the experiment of Fig. 2, we measure a photocarrier generation of the same order of magnitude as the typical doping level of the waveguide, and thus not sufficient to significantly modify the mobility.

*Variations of carrier density (ΔN).* In conventional silicon waveguides, free carrier density varies in presence of two-photon absorption (TPA) and surface-state absorption (SSA). Photocarrier generation by TPA increases quadratically with the local optical power $P$[38]

$$\Delta N = \frac{\tau_{eff} \beta_{TPA}}{2h\nu} \frac{P^2}{A_{eff}^2} \tag{s6}$$

where $\tau_{eff}$ is the carrier recombination time, $\beta_{TPA}$ is the TPA coefficient, $h\nu$ is the photon energy and $A_{eff}$ the waveguide effective area. Since $\Delta G$ increases linearly with $\Delta N$ (see Equations s3 and s4), TPA would require a quadratic relation between $\Delta G$ and $P$. As in Fig. 2 we do not observe such quadratic dependence, we rule out TPA as the physical mechanism responsible for the measured photocarrier generation. Instead, we measure sub-linear behavior, still in agreement with SSA, as pointed out in Section "Device impedance modeling and electrode design" of the Supplementary Information.

*Effective index and phase perturbation (derivation of Fig. 4b)*

The perturbation of the light signal measured at the output port of the resonator (Fig. 4a, purple line) is the result of a phase-to-amplitude modulation conversion given by the resonator spectral response[20]. The relative perturbation at the frequency $f_e$ of the electric signal driving the CLIPP can be expressed as

$$\frac{P(f_e)}{P_{CW}} = \frac{\partial H}{\partial \lambda} d\lambda = \lambda_r \frac{\partial H}{\partial \lambda} \frac{dn_{eff}}{n_g} \frac{L_e}{L_r} \tag{s1}$$

where $H$ is the intensity transmission spectrum of the ring resonator at the Through port, $d\lambda$ is the shift of the resonant wavelength $\lambda_r$ due to the induced perturbation $dn_{eff}$ of the waveguide effective index, and $n_g$ is the waveguide effective group index. The ratio $L_e/L_r$ takes into account that the index perturbation does not occur along the entire ring length $L_r$, but only in a portion of length $L_e$ between the two electrodes of the CLIPP, where the voltage $V_e$ is applied. In the example of Fig. 3a ($V_e$=1V) we measured a maximum relative perturbation $P(f_e)/P_{CW}= 1.65\cdot 10^{-3}$ (-27.8 dB) at the wavelength $\lambda_p=\lambda_r+9$ pm = 1555.898 nm, where $\partial H/\partial\lambda = 30$ nm$^{-1}$, corresponding to a resonance shift $d\lambda = 55$ fm (6.8 MHz). By substituting in eq. (s1) the values of the geometric ($L_e = 200$ μm, $L_r = 664$ μm) and optical ($n_g = 4.22$) parameters of the considered structure, we derived an effective index perturbation $dn_{eff} = 5\cdot 10^{-7}$, amounting to about 0.2 ppm of the waveguide effective index $n_{eff} = 2.45$. The corresponding phase modulation induced by the CLIPP is $d\phi = \dfrac{2\pi}{\lambda_p} dn_{eff} L_e = 0.4$ mrad. Considering a thermooptic coefficient of $1.84\cdot 10^{-3}$ K$^{-1}$ for the silicon waveguide[39], the induced perturbation is comparable to that induced by a temperature change of 2.7 mK.

For a better understanding of the practical implications of such a tiny perturbation, let us consider that a resonant frequency shift of less than 10% of the resonator linewidth can be considered acceptable in most applications. This implies that the CLIPP can be used to monitor resonators with a bandwidth as narrow as 50 MHz, that is with a Q factor of about $4\cdot 10^6$. Therefore, the CLIPP can be applied without any disturbance to the silicon microring resonators with the highest Q factor realized so far[21], which is in the order of $7.6 \cdot 10^5$.

For a better understanding of the physical mechanism originating such a small perturbation, we explored the behavior of the CLIPP up to an applied voltage of $V_e = 10$ V (see Fig. S5). The linearity of the measured perturbation versus $V_e$ rules out the existence of a significant thermal heating of the waveguide due to a current flow along the silicon core, which would rather exhibit a quadratic behavior versus $V_e$. Results are instead consistent with a linear electro-optic Pockels effect, according to which the variation $dn$ of the refractive index of the waveguide silicon core is related to the applied electric field by

$$dn = \frac{n^3}{2} rE = \frac{\chi^{(2)}}{n} E \qquad (s2)$$

where $\chi^{(2)}$ is the second order susceptibility. In our experiment, the change of the waveguide effective index $dn_{eff} = 5\cdot10^{-7}$ at $V_e = 1$V corresponds to $dn = 7\cdot10^{-7}$ and the average electric field in the 3-µm-thick silica layer underneath the electrode is $E = 0.3$ V/µm. These values lead to an electro-optic coefficient $r = 0.1$ pm/V ($\chi^{(2)} = 5$ pm/V) that is consistent with the values reported in the literature for non intentionally strained silicon waveguides[22].

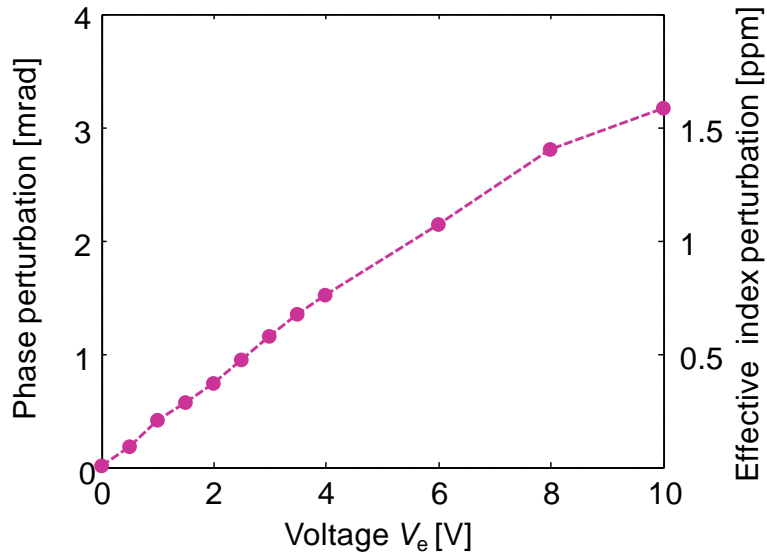

**Figure S5.** Experimental demonstration of the linearity of the measured phase and effective index perturbation induced by the CLIPP versus the applied voltage $V_e$.

**Estimation of surface carrier generation.**

The linear density of free carrier generated by SSA (Fig. 2, right vertical axis) is estimated by assuming that the density of photogenerated holes equals the density of photogenerated electrons, $\Delta N_h = \Delta N_e = \Delta N/2$, and that the mobility $\mu_{e,s} = 500$ cm$^2$/V·s and $\mu_{h,s} = 150$ cm$^2$/V·s of the carriers generated at the surface is about 30%[40] of the free carriers mobility in the bulk. Under these assumptions, Eqs. s3 and s4 lead to the following expression,

$$\Delta G = \Delta\sigma \frac{A}{L} = q(\frac{\mu_{e,s} + \mu_{h,s}}{2})\frac{A}{L}\Delta N, \qquad (s7)$$

from which the value of the linear density of surface free carriers ($A\Delta N$) plotted in Fig. 2 is calculated.